# Force spectroscopy on DNA by FM-AFM

Giovanni Di Santo, Susana Tobenas, Jozef Adamcik and Giovanni Dietler

Laboratoire de Physique de la Matière Vivante, Ecole Polytechnique Fédérale de Lausanne (EPFL), CH-1015 Lausanne

ABSTRACT We present imaging and force spectroscopy measurements of DNA molecules adsorbed on functionalized mica. By means of Non-Contact mode AFM (NC-AFM) in Ultra High Vacuum (UHV), the frequency shift ($\Delta f$) versus separation ($z$) curves were measured providing a quantitative measurement of both force and energy of the tip-DNA interaction. Similarly, topographic images of the adsorbed DNA molecules in constant frequency shift mode were collected. The high resolution force measurements confirm the imaging contrast difference between the substrate and DNA. The force curves measured along the DNA molecule can be divided into two classes showing marked differences in the minimum of the interaction force and energy, indicating that NC-AFM could deliver chemical contrast along the DNA molecule.

E-mail: giovanni.disanto@epfl.ch

INTRODUCTION

The structure of the DNA molecule was determined by means of X-ray diffraction already more than 50 years ago. Presently, its structure, its mechanical properties, its biochemical and biophysical properties and its biological activity are still under intense scrutiny. Single molecule techniques have opened new important possibilities. AFM is one of the single molecule techniques tuned toward structure determination [1] and interaction studies [2-3]. Notwithstanding that the present resolution has permitted only partial results on soft samples, topography images of DNA with increasing quality have demonstrate the capabilities of this technique to directly investigate the molecule structure [4-9]. Efforts were also devoted to study chromatin and its structure [11]. The reasons of the limited molecular resolution of AFM on DNA compared to the case of hard samples [12; 13] are attributed at least to five factors: (i) at room temperature, the thermal agitation of biomolecules is still important and unavoidable, this is related to need of flexibility and conformation fluctuations in order to carry out the biological activity; (ii) contact or intermitted contact AFM imaging are still the most used imaging modes, with the drawback of large interaction forces between tip and sample inducing deformation of the soft biomolecules; (iii) instrumental noise, related mostly to the thermal fluctuations of the cantilever, is still important because of the lever's low spring constant needed in order to limit the interaction forces and (iv) the tip-sample convolution mainly determined by the tip apex radius; (v) presence of a water and/or salt layer when imaging in air.

Recently, major advances were achieved pertaining to the combination of chemical contrast with spatial atomic resolution using NC-AFM on hard samples [12-15]. In these experiments stiff cantilevers were used but retaining low interaction forces. This raises the hope that, also on soft samples, NC-AFM can bring the needed sensitivity combined with a low interaction force necessary to reach nanometer or subnanometer resolution. Moreover, NC-AFM was also extended to liquid imaging with low force and high spatial resolution [16-18]. The aim of this communication is to relate the spatial resolution of the AFM to single point spectroscopy measurements done on DNA molecules deposited on APTES modified mica surfaces. The interaction force and potential differences on the substrate and along the DNA molecules were evaluated by direct calculation from the experimental data. These measurements provide also a realistic parameterization for the tip-molecule interaction to be used in further detailed theoretical modeling.

EXPERIMENTAL

All measurements were performed using a home built AFM operating at room temperature and under UHV conditions in Frequency Modulation (FM) [7; 19]. The operating parameters used in FM experiments were 4.5 to 40 N/m for the cantilever spring constant, resonance frequencies in 100-400 kHz range and free oscillation amplitudes of approximately 25 *nm*. Frequency shift $\Delta f$ versus distance *z* curves have been recorded with the necessary statistics (> 10 curves per point) both on the substrate and on the molecules. The force and potential curves were directly calculated from the raw $\Delta f$ vs *z* data by means of the method presented by Sader and Jarvis [20]. Topography images at constant frequency shift $\Delta f$ of single DNA molecules were recorded by means of NC-AFM operated at constant oscillation amplitude. The usual frequency shift used during imaging ranged from -16 *Hz* to -34 *Hz*. The AFM probes were Al-coated rectangular shaped Si cantilevers (MikroMasch Co.). The spring constant *k* was calculated using frequency scaling [21] that gave *k* values in good agreement within the nominal spring constant ranges. Circular DNA (plasmid *pBR322*, *Fermentas*) was diluted in TE (10 mM Tris-HCl, 1 mM EDTA, pH 7.8) to a final concentration of 1 mg/ml. The DNA plasmids were deposited onto the APTES modified mica according to our procedure described in [22] and immediately transferred in the UHV chamber. Low DNA concentration on the surface allowed visualization and force spectroscopy experiments on single DNA molecules.

RESULTS AND DISCUSSION

Results

Topography images of single *pBR322* plasmid DNA molecules were recorded by NC-AFM and are presented in Fig. 1. The AFM was operated in UHV and the images were taken at constant frequency shift. 90% of the deposited molecules present several crossings as a consequence of their supercoiled state (inset in Fig. 1*a*). Relaxed molecules, which are preferred if detailed topography and force spectroscopy measurements are aimed, appeared in the remaining 10% of the cases (Fig. 1*a* and 1*b*). The latter are either nicked DNA plasmids or truly not supercoiled molecules. The molecular sizes (averaged over 50 molecules) measured in the lateral ($7 \pm 1$ *nm*) and the vertical ($1.0 \pm 0.1$ *nm*) directions are in good agreement with similar measurements reported in the literature [5; 6; 23]. Experimental frequency shift $\Delta f$ versus distance curve sets are shown in Fig. 2. The interaction range of the forces (few *nm*) is wider than that presented by other authors on hard substrates [2; 12]. This can be related to the insulating character of the sample. Due to the polyelectrolyte character of the DNA molecule and of the mica surface, a contribution to the interaction force from electrostatic interaction (long range forces) is possible. The free oscillation amplitude (25 *nm*) used for these experiments enhances the contribution of those forces on the measured frequency shift. Besides that, the spectroscopic results contain detailed information about the tip-sample interaction. The comparison of the two sets of curves evidences a difference between the functionalized surface and the DNA molecule for the minimal value of the frequency shift $\Delta f$ of about 10 to 20 *Hz* combined with a *z* position displaced of approximately 0.5 to 1 *nm*. From these measurements one can also determine the contrast between substrate and DNA molecule measured in imaging mode (Fig. 2*b*), because the frequency shift signal is used as feedback for the *z* displacement control. From Fig. 2*a*, if a set frequency shift of $\Delta f$=-20 *Hz* is used in the experiments, one can extract from the intercepts between a horizontal line at $\Delta f$ =-20 *Hz* and the measured frequency shift curves, the "height" of the DNA molecule of approximately 0.5 to 1 nm. This is consistent with the average measured molecule height in respect to the substrate from images like in Fig. 1.

By analyzing the sets taken onto the DNA molecules (Fig. 2*a*; red and green dots), one can assign the force curves to two classes with a difference in frequency shift of ($5 \pm 1$) *Hz* for its minimal value while displaced of approximately ($300 \pm 50$) pm.

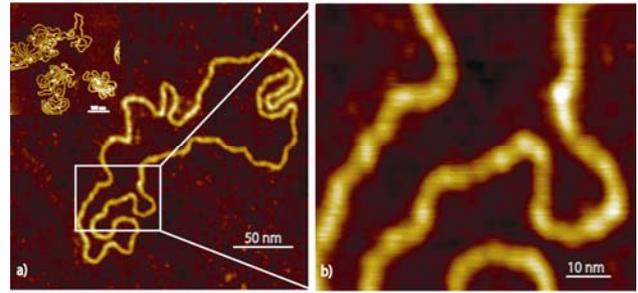

**FIG. 1** The images show *pBR322* plasmid DNA molecules deposited on APTES functionalized mica. Supercoiled DNA molecules with a high number of crossings are shown in the inset in (a). A typical relaxed molecule is shown in (a); detail in (b). Images and spectroscopy measurements have been recorded with the UHV-AFM working in NC mode (constant amplitude). Cantilever spring constant: *k*=14 *N/m*; Resonance frequency: $f_{res}$=294 *kHz* and frequency shift $\Delta f$ in the range between -16 *Hz* and -34 *Hz*; Free oscillation amplitude *A*=25 *nm*. Standard *Si* etched probe.

A possible explanation for this behavior is that the two classes correspond to the major and minor grooves of the double helix, respectively. The geometry allows us to speculate that in the case of the major groove (green), the tip is entering more inside the helical structure of the molecule whilst for the minor groove (red), the tip is entering less inside the DNA's double helix. This can explain why the probe senses a repulsive field $\Delta f$ >0) few hundreds of *pm* before. Furthermore, the difference in the minimal value of the frequency shift of the curves belonging to the two classes could be accounted for by the different chemical species interacting with the tip. In fact, the two grooves have a slightly different hydrophilic-hydrophobic character [24]. In order to quantify the measured frequency shift curves $\Delta f$ (*z*), force *F(z)* and potential *U(z)* curves were calculated using the method by Sader et al. [20]. The comparison between the force curves on DNA and on the substrate evidences a difference, at the minimum of the force curves, of approximately 350 *pN* (see Fig. 2*b*). As it is suggested in a recent publication, the minimum position in the force curve is a direct measure of the tip adhesion to the sample surface [25]. In this case we measure adhesion forces of -0.8 *nN* for APTES-mica while for the DNA is found at -0.45 *nN*. These smaller values in respect to the measurements shown by Schmutz et al. [25] are probably due to the reduction of the water layer in our UHV experiments.

The potential calculations returned smaller values for the potential curve minima on DNA ($\approx 7$ *eV*) compared with the curves on APTES-mica ($\approx 10$ *eV*) as it can be seen in Fig. 2*c*. In both plots of force and potential curves it is evident the difference between the substrate and the molecule in terms of force and potential as well as of *z*-shift.

Similarly, when the two classes of curves for the measurements on the DNA molecules are plotted together,

a small difference in interaction along the DNA is also evident and possible to quantify ($\Delta f \approx 100$ pN and $\Delta U \approx 1$ eV).

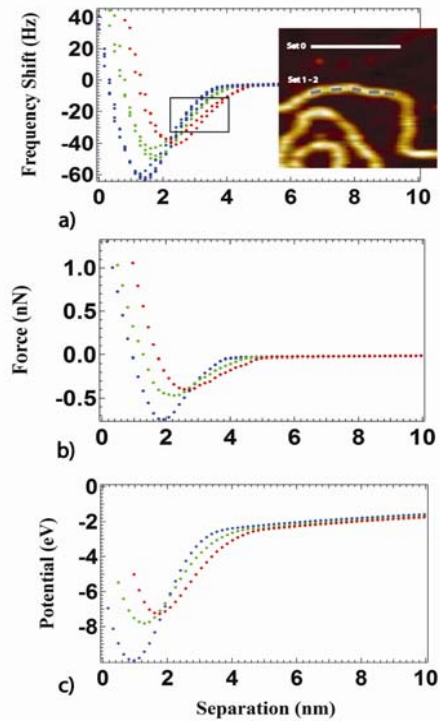

**FIG. 2** Frequency shift (a), force (b) and potential (c) curves on DNA/APTES-mica samples (blue: substrate; red and green: DNA molecule). The imaging operating set point is marked by the square in (a); optimal topography contrast was found for frequency shift values between -16 and -35 Hz. The calculated forces and potential curves show a difference between the substrate and the molecule of approximately 0.35 *nN* and 4 *eV* respectively.

The different hydrophilicity-hydrophobicity of the two grooves as well the chemical composition of the sugar-phosphate backbone should be very likely taken into account to explain this latter interaction energy difference. A second difference can be noted on the frequency force curves of Fig. 2*b*, namely the presence of longer range forces when the interaction force is measured above the DNA molecule compared to the case on the substrate. In fact in the region between 3 *nm* and 5 *nm* of distance, the slopes of the force curves for DNA and for the substrate are very much different.

Conclusions

NC-AFM images of DNA plasmids as well as frequency shift $\Delta f$ vs. distance $z$ curves were recorded with in UHV. The DNA images show a good spatial resolution and a contrast comparable to air experiments. From the frequency shift $\Delta f$ vs. distance $z$ curves we could explain the origin of the contrast between substrate and DNA molecule. Additionally, along one DNA molecule, two classes of frequency shift $\Delta f$, force $F(z)$ and potential $U(z)$ vs. distance $z$ curves were ascertained. The latter finding could indicate that the chemical composition of the minor-major groove and/or of the sugar-phosphate backbone along the molecule can be detected.